\documentclass[preprint]{aastex}
\usepackage{xspace}

\newcommand{\andi}{And\,I\xspace}
\newcommand{\andii}{And\,II\xspace}
\newcommand{\andiii}{And\,III\xspace}
\newcommand{\andv}{And\,V\xspace}
\newcommand{\andvi}{And\,VI\xspace}
\newcommand{\andvii}{And\,VII\xspace}
\newcommand{\ncl}{$\rm{N}_{\rm{C,L}}$\xspace}

\shorttitle{Survey for Carbon Stars in the M\,31 and Cetus dSphs}

\shortauthors{Harbeck, Gallagher, \& Grebel}

\begin{document}


\title{WIYN Survey for Carbon Stars in the M\,31 and 
 Cetus Dwarf Spheroidal Galaxies: Evolutionary Implications}


\author{Daniel Harbeck} 
\affil{Department of Astronomy, University of Wisconsin, 475 North Charter Street,
  Madison, WI 53706}
\email{harbeck@astro.wisc.edu}

\author{John S. Gallagher, III}
\affil{Department of Astronomy, University of Wisconsin, 475 North Charter Street, 
  Madison, WI 53706}
\email{jsg@astro.wisc.edu}

\and

\author{Eva K. Grebel}
\affil{Astronomical Institute of the University of Basel, Venusstrasse 7,
CH-4102 Binningen, Switzerland}
\email{grebel@astro.unibas.ch}



\begin{abstract}
We report results of a photometric survey with the WIYN telescope for
carbon stars in the M\,31 dwarf spheroidal (dSph) companions \andiii,
\andv, \andvi, and \andvii, as well as in the relatively isolated
Local Group dSph Cetus. We find three carbon-star candidates in
\andvii and one carbon star in each \andvi and Cetus. Comparing the
carbon star content with other Local Group dwarf galaxies, we argue
against the presence of substantial intermediate-age stellar
populations in the all of the galaxies surveyed with the exception of
\andvii.  We discuss these results in the context of the origin of the
Andromeda dSphs and conclude that these are ancient galaxies, most of
which ceased star formation long before the main merger events in
M\,31. The M\,31\, dSphs therefore show less diversity in star formation
histories than the Galactic dSph companions, or the M\,31 dE
companions, as illustrated by NGC~147 which was surveyed as a
calibration object.  All of our dSph targets except \andv have
candidate carbon stars below the tip of the RGB, which resemble CH
stars found in globular clusters.  We estimate that 0.3\% of stars in
the dSphs are CH stars, presumably as a result of C pollution from a
binary companion.  Comparisons with CH star frequencies in globular
clusters could constrain the impact of dense environments on the
frequency of this form of binary star evolution.

\end{abstract}




\section{Introduction}

The two massive spiral galaxies of the Local group (LG), the Milky Way
and Andromeda (M\,31) appear to have different histories.  While tidal
streams associated with minor mergers orbit both spirals (e.g.,
\citealt{ibata1995,ibata1997,ibata2001, ferguson2002, newberg2002,
  newberg2003, majewski2003, yanny2003}), the differing levels of
interactions in the histories of M\,31 and the Milky Way may be
reflected in the mean metallicity of their halo stars.  M\,31's halo is
more metal-rich than that of the Milky Way, suggesting that mergers
have had a substantial impact on M\,31 (see \citealt{durrell2001,
  brown2003} and references therein).

We also can ask to what degree the different formation histories of
M\,31 and the Milky Way are reflected in their systems of satellite
galaxies.  The dwarf spheroidal (dSph) galaxies of both systems are of
special interest because these galaxies might be the unused basic
building blocks in hierarchical galaxy formation scenarios (e.g.,
\citealt{bullock2001}), or, alternatively, what should have been
larger systems whose structure and evolution was modifed by their
proximity to giant spirals (e.g., \citealt{mayer2001}).  For a more
detailed discussion of this subject, see \citet{grebel2003}.

While deep photometric studies over wide fields for the Milky Way dSph
satellites reach below the oldest main sequence turnoffs (e.g.,
\citealt{hurley1999, harbeck2001, carrera2002, monelli2003, lee2003}),
allowing their star formation histories to be studied in detail, the
distance of the M\,31 dSph subsystem has so far limited existing
ground-based photometric studies to the upper red giant branch level
(e.g., \citealt{armandroff93,armandroff98, grebel99}) and space-based
photometric studies to the horizontal branch level (e.g.,
\citealt{dacosta2000,dacosta2002, harbeck2001, kara2003}). Thus, only
indirect conclusions can be derived for their star formation history
and mean age.  Even at this level the M\,31 dSphs appear to have unique
properties in that \citet{harbeck2001} found that the M\,31 dSphs show a
second parameter effect in their horizontal branch (HB)
morphologies. After accounting for offsets due to mean metallicities,
the M\,31 dSphs appear to have systematically redder HB morphologies
than the Milky Way's dSph companions. The origin of this second
parameter effect is not clear, but explanations include a slightly
(i.e., 2~Gyr) younger age of the M\,31 companions compared to their
Galactic equivalents (see \citealt{dacosta2000, harbeck2001}).  This
has also been suggested for Cetus (\citet{sarajedini02}).

In this paper we study the stellar populations in the M\,31 dSphs based
on properties of stars on the extended asymptotic giant branch (EAGB)
containing stars more luminous than the tip of the red giant branch
(RGB). These include classical EAGB carbon stars (C stars), which
trace intermediate-age-to-old (1 to 10 Gyr) stellar populations
(\citealt{aaronson1986}).  This class of C star results from thermal
pulses and subsequent mixing of freshly synthesized C into the
envelopes of the EAGB stars. The transition on the EAGB from an oxygen
rich (M star) to C star therefore depends both on the evolution and
initial abundances of the star; lower O abundances require less C to
make C/O$>$1.

A second class of C star is present even in old globular cluster
stellar populations with luminosities below the tip of the RGB. These
stars are products of mass transfer from a C star onto its main
sequence binary companion, which subsquently evolves up the RGB
(\citealt{dekool1995,han1995}).  The presence of C stars less luminous
than the tip of the RGB is evidence for binary evolution in a stellar
population, and is not tied to the age of the population in any simple
way. These are spectroscopically identified as CH stars; for a review
of CH and other types of carbon and related evolved stars, see
\citet{mcclure1985}. The CH stars in turn are likely to be evolved
versions of dwarf carbon (dC) stars (\citealt{dekool1995}), and we
discuss implications of the presence of CH stars in M\,31 dSph
companions in \S3.4.

While C stars are useful stellar population diagnostics, no systematic
wide-field survey for C stars in the M\,31 dSph galaxies has been
published so far. However, \citet{aaronson1985} found two EAGB C star
candidates in And~II, which led them to suggest that And~II sustained
star formation as an irregular-like galaxy for an extended time.
\citet{cote1999} spectroscopically confirmed an additional C star
located below the RGB tip luminosity, which they interpreted as being
a CH-like star, and probably a product of binary evolution of an
undetermined age. However, their color-magnitude diagram (CMD) of
And~II confirms the presence of an AGB and thus of an intermediate-age
stellar population. It also provides a good example of the importance
of spectroscopic confirmation in these small galaxies with few EAGB
star candidates (see their Fig. 4). In this paper we present a
systematic photometric survey using the WIYN 3.5-m
telescope\footnote{The WIYN Observatory is a joint facility of the
University of Wisconsin-Madison, Indiana University, Yale University,
and the National Optical Astronomy Observatories.} for C star
candidates in the Andromeda dSph companions \andiii, \andv,
\andvi, and \andvii\footnote{Note that \andvi and \andvii were named
Pegasus dSph and Cassiopeia dSph by their discoverers Karachentsev \&
Karachentseva (1999) after their parent constellations (see
\citet{grebel99} for their identification as dSphs and the
confirmation of their likely association with M\,31).  Peg dSph, which
was independently discovered by \citet{armandroff1999}, was named
\andvi by this group.  For simplicity, we will use the latter naming
convention for the M\,31 dSph companions in the following.}, and in the
comparatively isolated LG dSph galaxy Cetus. New data for NGC\,147 as
a calibrator for our survey technique will be presented as well.

\section{Data and Reduction}

We observed the Andromeda companions \andiii, \andv, \andvi, and
\andvii, as well as the Cetus dSph, with the MiniMo Mosaic CCD camera
at the 3.5m WIYN telescope located at Kitt Peak. We used Johnson V and
I filters to obtain temperature and luminosity information for stars
in the galaxies in order to construct a color magnitude diagram (CMD),
while observations in two narrow-band filters centered on the TiO and
CN feature at 778nm and 808nm, respectively, were used to identify
cool giant stars with enhanced carbon abundance. The use of a CN-TiO
filter combination is a well-established method for the reliable
identification of C stars (see \citealt{cook1986, albert2000,
nowotny2003}).  Most of the galaxies were observed during an observing
campaign on 2003 September 28 to 2003 October 1.

Since all galaxies are at a similar distance, the exposure times were
chosen to be 500\,s in each of the V and I filter, and 3$\times$500\,s
in each of the narrow band filters. There are two exceptions: The
exposure times in the V and I band filters for And VII were split into
2$\times$300\,s to avoid CCD blooming of bright stars present in the
field of view. The narrow band exposures for And VI were obtained
during an earlier observing run at WIYN on 2002 October 20; the
exposure times are 2$\times$500\,s per filter.  In addition we
observed the dwarf elliptical galaxy NGC\,147 to validate our ability
to identify C stars. This galaxy is known to have a large number of C
stars (approximately 146 C stars have been found in the work of
\citealt{nowotny2003}).

Although not always of photometric quality, the data are of superb
quality in seeing. The typical seeing in the near infrared narrow-band
filters was $0.5''$ to $0.6''$ and better; the typical seeing in the
broad band filters was $0.7''$, and never worse then $0.8''$. MiniMo
at WIYN consists of a mosaic of two CCD chips with a total field of
view of $9.6'\times9.6'$. The M\,31 dSph satellites with typical core
radii of order of $1'$ to $1.5'$ thus conveniently fit onto a single
chip, and have been centered on chip \#1 of the mosaic. The more
extended galaxies NGC\,147 and Cetus were placed at the center of the
field of view of MiniMo.

During the 2003 campaign there was a problem with the MiniMo CCD
readout electronics, resulting in strong random variations of the
CCD's overscan level. Before correcting the raw CCD frames with daily
zero calibrations frames, we subtracted the overscan on a line-by-line
basis. The variations in the overscan did not completely vanish,
leaving residual additional background noise with an RMS of order of
$5$ counts. Finally the images were corrected for illumination
variations using dome flat fields.

\begin{figure*}
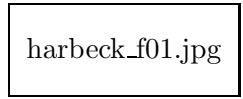

%\plotone{harbeck_f01.ps}
\begin{center}
\fbox{\rule[-4mm]{0cm}{1cm} harbeck\_f01.jpg }
\end{center}
\caption{$3.5\times3.5$\,' cut-out of the NGC 147 CN-filter
	 image (seeing $\le0.6$''). The box indicates the zoom area of
	 Fig.~\ref{fig_fits147zoom}, where we show an example of a
	 difference image. East is right, north down.}

\label{fig_fits147}

\end{figure*}

For each galaxy, all its images were registered to a common reference
coordinate system. The three (or two in the case of And VI) exposures
per narrow-band filter were stacked and used for cosmic-ray
rejection. In the case of \andvii we also stacked the two V and I
images. As an example we present a portion of the northern region of
NGC\,147 in Figure~\ref{fig_fits147}. Point spread photometry was
performed in all filters using the daophot implementation under
IRAF\footnote{IRAF is distributed by the National Optical Astronomy
Observatories, which are operated by the Association of Universities
for Research in Astronomy, Inc., under cooperative agreement with the
National Science Foundation.}, resulting in photometric catalogs for
each galaxy containing instrumental $V$, $I$, {\em TiO}, and {\em CN}
filter magnitudes.

Since the weather was non-photometric during parts of our observing
run, we did not attempt to observe photometric standard stars, and all
magnitudes cited in this paper are instrumental magnitudes. In order
to identify C stars, we only need differential photometry.  For
clarity, however, we adopted $I$, and $V-I$ zeropoints for each galaxy
to match the magnitude and color of the blue side of the tip of the
RGB with observed values (from \citealt{sarajedini02} for Cetus, from
\citealt{grebel99} for \andvii and \andvi, from \citealt{armandroff98}
and \citealt{armandroff93} for \andv and \andiii, respectively;
NGC\,147 is calibrated according to \citealt{nowotny2003}). We
calibrated the zeropoint of the {\em CN $-$ TiO} color using the
assumption that blue stars (i.e., bluer then the tip of the RGB)
should have {\em CN $-$ TiO} $=0$, since spectra of warmer stars are
expected to featureless in this wavelength region \citep{brewer1995}.

\subsection{Identification of Carbon stars}

The color-magnitude diagram for our test case, NGC\,147, is shown in
Fig.~\ref{fig_ngc147} (left). The intrinsically broad red giant branch
(RGB) features prominently as it does in the WFPC2 observations of
\citet{han1997}.  We also find a prominent population of stars well
above the tip of the RGB (TRGB). The right side of the same figure
demonstrates the selection of carbon stars from the two-color ($V-I$
vs.\ {\em CN $-$ TiO}) diagram: stars with high carbon content (and
therefore low TiO but strong CN molecule band absorption) stand out in
this plot and are selected according to the selection box ({\em CN $-$
TiO} $\ge$ 0.25). Stars in this plot with {\em CN $-$ TiO}
$\ge0.65$\,mag are almost always false detections, and are rejected
for NGC 147 (see discussion of false detections at the end of this
section). Only stars with $I\le21$\,mag are plotted in our two-color
diagram to limit this sample to objects with good photometry. Only
stars with colors of the RGB and $I\le20.7$\,mag (\citealt{han1997})
qualify as bona-fide C stars; stars fainter than the TRGB might be CH
stars and thus are not intermediate age stellar population tracers.  C
stars selected this way are plotted in the CMD of NGC\,147 with filled
circles. The substantial number of C star candidates (155) matches our
expectation for this galaxy and compares well to the 146 C stars found
by \citet{nowotny2003}.

In addition to the photometric selection, we construct the difference
of the CCD images observed in the TiO and CN filters to cross-identify
the candidates on these images. An example of such a difference image
is given in Fig.~\ref{fig_fits147zoom}, where we zoom into the boxed
region of Fig.~\ref{fig_fits147}.  A visual inspection of the
difference image of NGC\,147 leads us to estimate a roughly 10\%
incompleteness in our C star detections, which is compensated by
roughly 10\% false detections. We did not cross-identify each star in
our NGC\,147 catalog with the difference image, but will do so for the
dSph galaxies and will accept only C stars that appear both in our
photometric catalog and stand out as a clear detection in the
difference images. The fact that we find a slightly larger number of C
stars in NGC\,147 than \citet{nowotny2003} can be explained by our
deeper photometry, better seeing conditions (by at least $0.3''$ in
the narrow-band filters), and a larger field coverage.  A comparison
between the catalog of C stars of \citet{nowotny2003} indeed reveals
that most of the stars are in common with our catalog, with both false
detections and misses of order of 10\% in their as well as in our
work. We thus increase the number of C stars in NGC\,147 to
approximately 155, and conclude that our selection efficiency of C
star candidates is comparable to other surveys.

Due to the small absolute number and an improved crowding situation,
we slightly loosen the photometric selection criterion for C stars in
the dSphs, by requiring a {\em CN $-$ TiO} color of $\ge0.2$\,mag and
a luminosity of the C star candidates brighter than $I=22$\,mag. This
selection will result in some false detections that we identified and
removed during the cross-correlation with the difference image. With
the small number of C star candidates per galaxy we can afford to
double-check each individual star. Reasons why stars might appear in
the photometric catalog and not in the difference image turned out to
be (and are indeed limited to): misidentification of background
galaxies, of cosmic ray trails (not all cosmic rays are actually
removed), stars affected by stray light or blooming in the vicinity of
bright stars in the field of view, blends of two stars, or CCD
defects. C star candidates that turned out to be false detections have
been marked with open circles in the diagrams. Stars that remain valid
C star candidates after cross-identification are plotted with filled
circles.

\subsubsection{The Andromeda dSphs}

The CMDs and two-color diagrams of the observed M\,31 companions are
presented in the same manner as for NGC\,147 in Figs.\,\ref{fig_and3}
to \ref{fig_and7}. The color of the RGB depends on the metallicity of
the galaxies, and we carefully adjusted the $V-I$ color boundary for
the C star selection to avoid confusion with stars bluer than the tip
of the RGB. The initial number of photometric detections of C star
candidates that are {\it not} cross-identified in the difference image
is very different for the four observed galaxies (as indicated by the
number of open circles in the diagrams), and turned out to be clearly
correlated with the number of (bright) field stars on the CCD frame.
The larger the number of bright foreground stars, the larger the
number of false detections.  After comparing the initial photometric C
star candidates with the difference images, we are left with a smaller
set of reliable C star candidates in the galaxies.  These are listed
in Table~\ref{tab_candidates}.

Depending on whether the bolometric luminosity of a C star candidate
is brighter or clearly dimmer than the tip of the RGB, we classify it
as a genuine C star (``C'' in Table~\ref{tab_candidates}) or as an
evolved dC star (also called ``CH'' star), respectively. This
selection criterion is unique for all but one star in \andvii (see
Figure\,~\ref{fig_and7}), which has almost the same luminosity as the
TRGB. We classify this star as a C star candidate.  This is justified
as technically AGB stars would have at least the same bolometric
luminosity as the tip of the RGB. If we apply a correction in the form
$B.C. = -0.246 \cdot (V-I)$ \citep{reid1984} this star would clearly
satisfy the bolometric luminosity criterion.

The V- and I-band images and the TiO- and CN-band images of \andvi
were taken during two different campaigns, and due to imperfect
overlap of the two observations the total field covered of \andvi is
limited to a $2.9'$-wide stripe (more than twice the core radius of
$1.3'$, \citealt{caldwell1999}). The number of detected C stars for
this galaxy therefore could be incomplete.

The lack of radial velocity information for the C star candidates
prohibits prevents us from verifying their membership in the M\,31 dSph
companion galaxies.  Owing to the large angular distance of these
galaxies from M\,31, confusion with M\,31 halo C stars seems unlikely.
Indeed, all of our C star candidates are near the central regions of
the dSphs, supporting the idea that younger (that is,
intermediate-age) populations are primarily found in the central
regions of dSphs (see \citealt{gre99,gre00, harbeck2001}).  The
centralized location of the C stars and their narrow band colors makes
it unlikely that we are seeing misidentified Galactic foreground dC
stars or backgrund galaxies.  In the remainder of this paper we will
assume that the C star candidates are members of the M\,31 dSph
satellites.

\subsubsection{Cetus}

Cetus was placed on the center of the MiniMo field of view, so there
is no photometry available for a central $7''$ wide stripe. The CMD
and the two-color diagram of Cetus are shown in
Figure~\ref{fig_cetus1} and Fig.~\ref{fig_cetus2} for the MiniMo CCD
chip\,1 and chip\,2, respectively. We can identify three objects with
enhanced carbon abundance.  One of them is a likely C star candidate,
while two of them appear to be CH stars or similar objects.  However,
due to the gap in the MiniMo image we might miss approximately one
($1\pm1$) C star in Cetus. In good agreement with earlier studies
(\citealt{whiting1999, sarajedini02}) we do not find clear evidence
for an EAGB in this galaxy.

\section{Discussion}

We successfully detected high probability candidates both for EAGB C
stars and CH stars in four M\,31 dSphs surveyed and the Cetus dSph. In
his reviews of the C star content of Local Group galaxies,
(\citealt{groene1999,groene2002}) plots absolute magnitude of LG
galaxies versus the logarithm of the number of C stars, carefully
corrected for area coverage.  We expanded this plot with new data
points as shown in Figure~\ref{fig_mvlnc} using new results from the
literature as summarized in Table~\ref{tab_cstats}.  Absolute $V$-band
magnitudes were taken from \citet{grebel2003}.

We consider only LG galaxies where surveys for C star candidates using
the CN-TiO narrow-band filter technique are expected to be reasonably
complete.  While several C stars have been found in Milky Way dSph
galaxies \citep{aaronsonmould1985, groene2002}, in some cases these
are likely to be CH-stars, and the surveys often have incomplete areal
coverage.  We therefore include only a few Milky Way dSph satellites
in Figure~\ref{fig_mvlnc}.  For the SMC and LMC we adopt the DENIS
infrared survey results for the number of C stars
\citep{cioni2003}.

The galaxies in Figure~\ref{fig_mvlnc} show the well-known correlation
between the absolute V-band luminosity of a galaxy (i.e., an
approximate stellar mass) and the total number of C stars.  A useful
tool to study the C star content of a galaxy is \ncl$=\log(\#C) + 0.4
\cdot \rm{M}_{\rm{V}}$, which measures the number of C stars per unit
luminosity, and thus is a measure for the deviation from the
M$_V$-log(\#C) relation (\citealt{aaronson1983, groene1999}).  We plot
\ncl vs.\ the mean metallicity of the galaxies in Fig.~\ref{fig_mvlnc}
(right).  \ncl scatters between -3 and -4.6 for the galaxies, as
previously illustrated by \citet{groene1999,groene2002}. There is no
apparent trend with metallicity, consistent with the predictions of
theoretical population models \citep{mouhcine03}.  These models show
that for smoothly evolving stellar populations, \ncl is expected to
drop substantially only for super-solar metallicities or systems where
the stars are all ancient with ages $\ge$9-10~Gyr.

We divide the galaxies in this sample into three classes, motivated by
their morphology and --- where known --- their star formation history:
(i) ``stellar fossil'' dSphs, where star formation ceased at very
early times, (ii) dwarf irregular galaxies, transition-type galaxies,
and dSphs where the period of star formation extends over a
substantial fraction of cosmic time, and (iii) the dE companions of
M\,31.  Of the Galactic dSphs included in Table~\ref{tab_cstats}, a
deep HST study of the stellar populations in Leo\,I revealed that this
galaxy was actively forming stars until 1\,Gyr ago, with slower
ongoing star formation until $\sim$300\,Myr before the present
\citep{caputo1999,gallart1999}. The situation is similar for Fornax,
where star formation stopped only $\sim$100 to 200\,Myr in the past
( \citealt{stetson1998, greste1999};  for recent reviews of the star
formation histories of Local Group dwarf galaxies, see
\citealt{gre99,gre00}). These two Galactic dSphs are in our second star
forming galaxy category.  All of the M\,31 dEs have large numbers of
intermediate age stars, and low level star formation continues into
the present in NGC~185 and NGC~205 (\citealt{lfm93, grebel97,
delgado99, capellari99, davidge2003}). We omit the compact elliptical
M32 from our discussion due to the lack of a published survey for C
stars.

The different symbols in Fig.~\ref{fig_mvlnc} refer to the three
classes of galaxies we just defined.  We can extract a general trend
from Fig.~\ref{fig_mvlnc} (right panel): Quiescent dSph galaxies
contain a small number of C stars per V-band luminosity ($\le$ 10$^{-4.5}$),
while those galaxies with active or recent star formation have a
higher C star content ($\approx 10^{-3.5}$). The recent star formation
in the Fornax dSph makes this an outstanding galaxy in this plot.  The
normalized C star content of the three M\,31 dEs (NGC\,147, NGC\,205,
and NGC\,185) scatters around N$_{C,L}=-4$, consistent with their
large intermediate-age stellar populations.

\subsection{Star Formation Histories of the Andromeda dSphs}

The CMDs derived from WFPC2 imaging of \andi, \andii, \andiii, \andv,
\andvi, and \andvii do not extend much below the horizontal
branch. They are therefore sensitive only to potential main sequence
stars that have ages of $\leq$1~Gyr. None of these dSphs show evidence
for star formation within this recent time frame (see, e.g.,
\citealt{harbeck2001, kara2003}).  Furthermore, the CMDs presented by
\citet{harbeck2001} show that red clumps, typical of intermediate age
stars, are unlikely to be present in the M\,31 dSphs, and that all the
M\,31 dSphs have pronounced horizontal branches.
 
\subsection{The C/M ratio}

Comparisons of the number of normal EAGB stars, seen as stars with
spectral classes of M5 or later, to C stars also is a useful
evolutionary diagnostic (see, e.g.,
\citealt{blanco1983,mouhcine03,cioni2003}).  This ratio depends on
both age and metallicity, with the relative number of C stars peaking
for mean ages of $\approx$1~Gyr, and in all cases declining for ages
greater than $\sim$10~Gyr. Unfortunately, the relevant ratio of
$N_C/N_{M5+}$ is very difficult to derive for the M\,31 dSphs due to the
very small number of C and M stars on the EAGB and the presence of
substantial numbers of red Galactic foreground stars.

An approximate value of $N_C/N_{M5+}$ can be obtained from our data by
counting all stars redder than $V-I\ge2.0$\,mag (to select $M5+$
spectral classes; \citealt{richer1985}, \citealt{pritchet1987}) and
$I<I_{\rm{TRGB}}$ (i.e., brighter than the tip of the RGB), or $I>19$.
However, we avoid stars near to the saturation limit of the CCD, which
are bright foreground stars, and objects whose peculiar colors
indicate either that they are not stars or that a problem exists with
the data. This approach is most effective when the density of EAGB
stars relative to foreground field stars is $>$1, as in NGC~147.

We counted all EAGB $M5+$ stellar candidates in NGC\,147 both in the
full field of view (which will lead to an overestimation of $N_{M5+}$
due to field contamination), and within the central $3'$, which may
cause us to slightly underestimate $N_{M5+}$ due to crowding
effects. This approach yields $N_C/N_{M5+} = \frac{159}{1024} = 0.16$
and $\frac{111}{575}=0.19$ for the full field and the central region,
respectively. Given the expected uncertainty in $N_C/N_{M5+}$ of order
of 0.02 we are in excellent agreement with the result of
\citet{nowotny2003}, who found $N_C/N_{M5+}=0.15$ for NGC~147.

Since the M\,31 dSphs cover a relatively small region of the WIYN MiniMo
camera CCD chips, we counted all stars fulfilling the conditions
defined above in strips above and below the dSph targets. The
resulting estimate of the foreground Galactic star contamination at
\andvi is too large relative to any population of EAGB $M5+$ stars to
get a meaningful estimate for $N_{M5+}$.  Similarly, we find no M-star
candidates in the Cetus dSph, where this method yields $N_{M5+} =
-$2$\pm4$.  We also see that since \andiii and \andv have no EAGB C
stars, they must have $N_C/N_{M5+}=$0.

For \andvii this method gives $N_{M5+} \approx48$.  We derive a
relative fraction of C stars of $N_C/N_{M5+} = 0.06$ or $\log(C/M)
\approx -1.2$.  For a galaxy with constant star formation we would
expect $\log(C/M)\ge 1$ at [Fe/H] $\approx -$1.5 (e.g., Fig.\,8 in
\citealt{mouhcine03}). If we adopt a correction for the bolometric
luminosity (as in \citealt{reid1984}) to select M stars, we would
slightly find 38 M5+ stars despite the fact that we would include even
redder stars with I-band luminosities fainter than the tip of the RGB
by the selection, which mainly seems to reflect the uncertainty in the
background subtraction.  Despite the considerable uncertainty in this
ratio, we can say there is no indication for the presence of a
significant younger intermediate-age ($<$3~Gyr) stellar population in
\andvii based on the models of \cite{mouhcine03}.

The presence of only one carbon star in \andvi and Cetus, where we
find no significant populations of EAGB $M5+$-stars might be
interpreted as a high $N_C/N_{M5+}$ ratio which would be unexpected
for a small young stellar population. However, with only one candidate
EAGB star in each galaxy, a range of possibilities existis, including
the production of an EAGB C star via binary evolution. For example, C
stars should be able to form from blue stragglers.  Furthermore, with
only one EAGB candidate per galaxy, spectroscopic confirmation of
membership is essential.

In summary, only \andvii in our sample shows evidence for a
substantial EAGB stellar population. In this case the low value of the
$N_C/N_{M5+}$ ratio places these stars in the medium range of
intermediate stellar ages, namely ages $\ge$3-5~Gyr.
 
\subsection{On the origin of the Andromeda dwarf companion galaxies}

We began this investigation with the objective of comparing the dSph
companions of M\,31 with those of the Milky Way, and of investigating
whether the dSphs might be linked to the merger history of M\,31. Our
data are consistent with the M\,31 dSphs having predominantly old
stellar populations, where most star formation occurred at $>$10~Gyr
in the past.  The only clear dSph exception are \andii (from 
\citealt{aaronson1985} work), and \andvii, where significant star formation
could have continued until 3-5~Gyr ago. Both the survey for EAGB stars
presented here and the study of horizontal branch morphologies by
\cite{harbeck2001} are consistent with the view that the majority of M\,31's
 dwarf satellites are old systems. In particular, we do not find
 examples of dSphs with recent star formation activity, like the
 Fornax or Carina dSph satellites of the Milky Way.

Any connections between the dSphs and stellar halo of M\,31 remain
tenuous. As \cite{ferguson2002} emphasize, even though \andi and
\andiii lie in the projected vicinity of the main M\,31 southwestern
stellar stream, their low metallicities make dSphs unlikely to have
been dominant contributors to the more metal rich components of the
M\,31 halo.  A further constraint comes from the \cite{brown2003}
analysis of their spectacularly deep {\it HST} CMD of the M\,31 stellar
halo. They find that the significant metal-rich halo component of
M\,31 has an age of 6-8~Gyr, too young to be associated with any of
the M\,31 dSph companions with the possible exception of the distant
\andvii dSph; see \citet{grebel2003} for deprojected distances and
mean metallicities of the M\,31 dwarf companions.

In terms of their star-formation histories, M\,31's dSphs resemble the
ancient Galactic dSph companions as epitomized by the Draco and Ursa
Minor dSphs.  Of these two, Draco appears to be dark-matter dominated
(e.g., \citealt{odenk2001}), whereas Ursa Minor may be undergoing
tidal disruption (e.g., \citealt{palma2003}), so Draco probably is the
best comparison object.  The M\,31 dSph companions evidently formed as
independent objects that are not directly associated with the recently
uncovered indications of interactions with M\,31, which took place
during the galactic ``middle ages'' some 6--8~Gyr in the past.

However, a possible environmental effect is seen in that only two
M\,31 dSphs evidently supported extended periods of star formation,
fewer than in the population of Galactic dSphs located at similar
distances from their spiral host galaxy.  That present-day environment
cannot be the only agent becomes apparent when considering the
similarly quiescent, isolated Cetus dSph (see \citet{grebel2003} for a
more extensive discussion).  An independent constraint on the
importance of dSphs for the accretion history of M\,31 could be
established by a study of the [$\alpha$/Fe] abundances in both M\,31 and
its companions to see whether halo and dSphs exhibit similar
differences as found for the Milky Way by, e.g.,
\citet{shetrone2001}. Unfortunately, because of the distance of M\,31
and its satellites only global [Fe/H] values can be measured
spectroscopically to date (\citealt{cote1999, guha2000, reitzel2002}).

\subsection{On the frequency of dC/CH stars}

The standard scenario of ``extrinsic'' C star formation implies that a
main sequence star accretes material through Roche-lobe overflow from
its intrinsic EAGB C star binary companion. This mechanism also is
responsible for CH-stars (see \citealt{mcclure1990},
\citealt{han1995}) and we can therefore consider the dC and CH-like
stars to be tracers of similar effects of binary evolution. But what
fraction of stars in a stellar population will suffer from pollution
severe enough to turn them into C stars? This number depends both on
the binary fraction, and the distribution of binary star semi-major
axes, metallicity, and age (see, e.g., \citealt{dekool1995}).  A
``hard'' binary system with a orbital separation smaller then a
typical AGB radius might disrupt the evolving AGB star before a third
dredge-up is able to turn the AGB star into a C star. Binary-orbit
shrinking due to close three-body encounters in dense globular
clusters could reduce the specific frequency of dC stars
\citep{cote1997}. Thus the frequency of dC and CH stars is expected to
depend also on the environment, and we should find relatively more
dC/CH stars in low density dSph galaxies with old stellar populations
than in globular clusters.

To estimate the relative number of CH-stars in the uncrowded dSph
galaxies, we counted the number of RGB stars brighter than the
$I=22$\,mag limit of our C star survey. The ratio between numbers of
CH-stars, which we approximate as evolved dC stars, and RGB stars is
an approximate measure of the relative fraction of dC stars in the
stellar populations of these galaxies. However, the small number
statistics of the rare C-rich stars introduces substantial
uncertainties into this ratio.  In Table~\ref{tab_dcfrequency} we
present the number of RGB stars and CH star candidates, as well as the
estimated percentage of CH stars for the galaxies surveyed. We assume
that all of our candidates are CH stars in the M\,31
dSphs. Spectroscopic follow-up of these stars is essential to confirm
their membership by their radial velocities and verify that they are
C-rich objects.

To improve the statistics, we simply sum up all of the RGB stars and
CH stars to obtain a first estimate of the dSph CH star frequency. We
find $\sim$0.3\% of the dSph's stars to be CH stars that could have
originated from dC stars which were contaminated during the evolution
of a binary stellar system.  The frequency of CH stars in globular
clusters remains to be established since such stars are very rare (see
\citealt{cote1997}).

In NGC\,147 we count CH stars in the same way as we did for the dSphs
between the tip of the RGB and $I\le22$, and also count RGB stars in
the same luminosity range. This way we derive a slightly higher CH
stars frequency of $0.8$\%, which might be unreliable since strong
crowding might become important for the selection of fainter
carbon-rich stars and confusion with genuine C stars at the bright end
of the selection might lead to an overestimation of the number of CH
stars.

\section{Summary}

We present results from a photometric survey for C stars and related
carbon-rich stars in the \andiii, \andv, \andvi, \andvii, and Cetus
dSph galaxies carried out with the WIYN 3.5-m telescope. Similar data
also were obtained for the dE galaxy NGC~147, previously surveyed for
C stars by \cite{nowotny2003}, as a check on our technique.  The small
number of C stars found among the M\,31 dSph satellite galaxies
implies that they contain predominantly old stellar populations. Aside
from \andii and especially \andvii, where star formation may have been
active as recently as 3-5~Gyr in the past, the \andiii, \andv, and
\andvi dSphs have not supported significant star formation within the
past $\sim$10~Gyr.

The stellar populations of the M\,31 dSphs both are older and more
metal-poor than the intermediate-age M\,31 stellar halo components. We
conclude that the dSph satellites are an early legacy of the time
before the prominent M\,31 merger events that occurred about 6-8~Gyr
in the past.  The lack of a connection to the M\,31 ``middle ages''
merger is further supported by the similarities in stellar populations
between the M\,31 satellites and relatively isolated Cetus dSph, as
well as the Galactic dSphs with old populations, such as the Draco or
UMi dSphs.

The frequency of C stars in the M\,31 dE companion NGC~147 from our
data and the recent study by \cite{nowotny2003}, places it at levels
the actively star forming Magellanic Clouds and Fornax or Leo~I,
Galactic dSphs where star formation continued to within $\sim$1~Gyr of
the present.  While the M\,31 dEs structurally resemble scaled up dSph
systems, they have very different star formation histories than most
of the M\,31 dSphs.  This serves to emphasize the profound differences
in stellar population characteristics among the M\,31 dSph
satellites. The low luminosity dSphs are predominantly stellar
fossils, while the dEs made many of their stars 5-10~Gyr in the
past. These dE galaxies therefore may be associated with some of the
more pronounced merger features in M\,31, as others have suggested on
the basis of their structures and mean stellar metallicities
(\citealt{ferguson2002}). The \andvii dSph is neither a dE-like object
nor a collection of primarily ancient stars, and instead resembles
Galactic dSphs which experienced star formation for more than
$\sim$5~Gyr after their formation.

We also found that roughly $0.3$\,\% of the stars in dSphs to be
candidate CH stars, products of binary evolution.  Future
investigations of dC/CH fractions in globular clusters will allow to
investigate the impact of a dense environment on binary evolution.

\acknowledgements

These observations reflect the dedicated efforts of the WIYN
Observatory staff, and we thank them for their support.  DH gratefully
acknowledges support as a McKinney postdoctoral fellow and from the
Graduate School at the University of Wisconsin-Madison.  JSG
acknowledges essential funding from NSF grant AST-9803018 to the
University of Wisconsin and expresses his appreciation to Andrew Cole,
Ariane Lan\c{c}on, and Mustapha Mouhcine for discussions regarding the
properties and importance of E-AGB stars.  EKG and JSG also thank the
Swiss National Science Foundation for partial support, and all three
authors express their appreciation to the Max-Planck-Gesellschaft for
supporting stimulating workshops on dwarf galaxies. This research has
made use of NASA's Astrophysics Data System (ADS) and the NASA/IPAC
Extragalactic Database (NED) which is operated by the Jet Propulsion
Laboratory, California Institute of Technology, under contract with
the National Aeronautics and Space Administration.

\clearpage

\clearpage
\begin{figure}
%\plotone{harbeck_f02.ps}
\begin{center}
\fbox{\rule[-4mm]{0cm}{1cm} harbeck\_f02.jpg }
\end{center}
\caption{Zoom ($\approx 45\times45$\,'') into a difference {\em  TiO$-$CN } 
          image in the region boxed in Fig.~\ref{fig_fits147}. C stars
          with weak TiO absorption and strong CN absorption appear as
          bright objects (encircled), while normal stars just subtract
          out.  The boxed region shows an example of imperfect image
          subtraction of brighter stars; these objects do not appear
          in the C star catalog thanks to a cut in the candidates
          luminosity.}
\label{fig_fits147zoom}
\end{figure}

\clearpage

\begin{figure*}
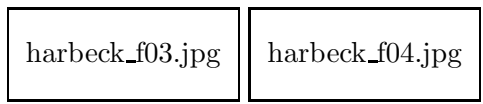

%\plottwo{harbeck_f03.ps}{harbeck_f04.ps}
\begin{center}
\fbox{\rule[-4mm]{0cm}{1cm} harbeck\_f03.jpg }
\fbox{\rule[-4mm]{0cm}{1cm} harbeck\_f04.jpg }
\end{center}
\caption{ Color-magnitude diagram (CMD; left) and two-color diagram (right)
  of NGC 147; all magnitudes are instrumental, but zeropoint corrected
  as described in the text. The two-color plot shows the broad-band
  colors of stars in NGC 147 (i.e., a temperature) versus the relative
  CN absorption strength as measured by the CN-TiO filter colors; only
  stars brighter than $I=22$\,mag are plotted (indicated by a line in
  the CMD, left). Red stars with strong CN absorption falling into the
  selection box are identified as Carbon stars and are clearly marked
  with filled circles in both plots.}
\label{fig_ngc147}
\end{figure*}

\clearpage

\begin{figure*}
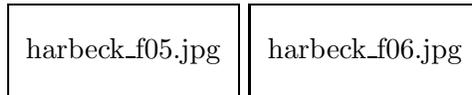

%\plottwo{harbeck_f05.ps}{harbeck_f06.ps}
\begin{center}
\fbox{\rule[-4mm]{0cm}{1cm} harbeck\_f05.jpg }
\fbox{\rule[-4mm]{0cm}{1cm} harbeck\_f06.jpg }
\end{center}
\caption{Same as Fig.~\ref{fig_ngc147}, but showing the CMD and
  two-color diagram for the And III dSph galaxy. C-star candidates
  that could be cross-identified on the TiO-CN difference image are
  plotted with filled circles, false detections with open circles. }
\label{fig_and3}
\end{figure*}

\clearpage

\begin{figure*}
%\plottwo{harbeck_f07.ps}{harbeck_f08.ps}
\begin{center}
\fbox{\rule[-4mm]{0cm}{1cm} harbeck\_f07.jpg }
\fbox{\rule[-4mm]{0cm}{1cm} harbeck\_f08.jpg }
\end{center}
\caption{Same as Fig.~\ref{fig_and3} for the And V dSph galaxy. }
\label{fig_and5}
\end{figure*}

\clearpage

\begin{figure*}
%\plottwo{harbeck_f09.ps}{harbeck_f10.ps}
\begin{center}
\fbox{\rule[-4mm]{0cm}{1cm} harbeck\_f09.jpg }
\fbox{\rule[-4mm]{0cm}{1cm} harbeck\_f10.jpg }
\end{center}
\caption{Same as Fig.~\ref{fig_and3} for the And VI dSph galaxy. }
\label{fig_and6}
\end{figure*}

\clearpage

\begin{figure*}
%\plottwo{harbeck_f11.ps}{harbeck_f12.ps}
\begin{center}
\fbox{\rule[-4mm]{0cm}{1cm} harbeck\_f11.jpg }
\fbox{\rule[-4mm]{0cm}{1cm} harbeck\_f12.jpg }
\end{center}
\caption{Same as Fig.~\ref{fig_and3} for the And VII dSph galaxy. }
\label{fig_and7}
\end{figure*}

\clearpage

\begin{figure*}
%\plottwo{harbeck_f13.ps}{harbeck_f14.ps}
\begin{center}
\fbox{\rule[-4mm]{0cm}{1cm} harbeck\_f13.jpg }
\fbox{\rule[-4mm]{0cm}{1cm} harbeck\_f14.jpg }
\end{center}
\caption{Same as Fig.~\ref{fig_and3} for stars of the Cetus dSph that
  fall on Chip \#1. }
\label{fig_cetus1}
\end{figure*}

\begin{figure*}
%\plottwo{harbeck_f15.ps}{harbeck_f16.ps}
\begin{center}
\fbox{\rule[-4mm]{0cm}{1cm} harbeck\_f15.jpg }
\fbox{\rule[-4mm]{0cm}{1cm} harbeck\_f16.jpg }
\end{center}
\caption{Same as Fig.~\ref{fig_and3}, but for stars of the Cetus dSph that
  fall on Chip \#2.}
\label{fig_cetus2}
\end{figure*}

\clearpage

\begin{figure}
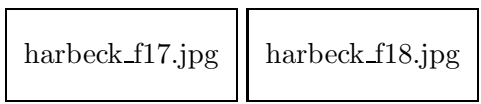

%\plottwo{harbeck_f17.ps}{harbeck_f18.ps}
\begin{center}
\fbox{\rule[-4mm]{0cm}{1cm} harbeck\_f17.jpg }
\fbox{\rule[-4mm]{0cm}{1cm} harbeck\_f18.jpg }
\end{center}
\caption{Left: The 
 relation between the number of C stars and the luminosity of Local
 Group dwarf galaxies. Right: The same as left, but with the mean
 metallicity on the X-axis.}
\label{fig_mvlnc}
\end{figure}

\clearpage

\begin{deluxetable}{llllll}
\tablewidth{0pt}
\tablecaption{C Stars in Local Group dSphs}

\tablehead{
 \colhead{Galaxy} & \colhead{Id} & \colhead{Ra\tablenotemark{a}} &
 \colhead{Dec\tablenotemark{a}} &
 \colhead{M$_{\rm{I}}$\tablenotemark{b}} &
 \colhead{dC/C\tablenotemark{c}} }

\startdata
And III & and3-2538  &  0:35:35.19 & 36:29:00.3 & 21.0    & dC      \\
And V   & \nodata    & \nodata     & \nodata    & \nodata & \nodata \\
And VI  & and6-6618  & 23:51:46.63 &  24:34:59.3 & 20.2    & C       \\
        & and6-10813 & 23:51:58.96 &  24:34:46.9 & 21.8    & CH      \\
And VII & and7-3727  & 23:26:30.78 &  50:40:50.6 & 20.7    & C        \\
        & and7-5653  & 23:26:37.54 &  50:41:28.5 & 21.0    & C        \\
        & and7-6111  & 23:26:39.56 &  50:39:49.1 & 20.9    & C        \\
        & and7-4002  & 23:26:31.70 &  50:40:57.6 & 21.3    & CH       \\
        & and7-4447  & 23:26:33.20 &  50:41:08.6 & 21.4    & CH       \\ \hline
Cetus   & cetus-II705 & 00:26:12.99 & -11:01:20.2 & 20.6   & C     \\
        & cetus-II736 & 00:26:13.64 & -11:02:16.7 & 21.3   & CH    \\
        & cetus-I3089 & 00:26:13.17 & -11:03:59.7 & 21.1   & CH    \\

\enddata 
\tablenotetext{a}{J2000.0, based on USNO2.0 catalog}
\tablenotetext{b}{Instrumental magnitudes, but zero point corrected in I.}
\tablenotetext{c}{Classification into genuine Carbon star (C) and
  evolved dwarf Carbon star (CH)}

\label{tab_candidates}
\end{deluxetable}

\clearpage

\begin{deluxetable}{lccccl}
\tablewidth{0pt}
\tablecaption{C stars in LG dwarf galaxies}

\tablehead{
 \colhead{Galaxy}                        &
 \colhead{M$_V$\tablenotemark{1} }  &
 \colhead{$\log{}$\#C}                     &
 \colhead{[Fe/H]\tablenotemark{1} }      &
 \colhead{Type\tablenotemark{1} }        &
 \colhead{Source\tablenotemark{2}} 
 }

\startdata 

Phoenix    &  -9.8 &  0.3     & -1.9 &   2 & 1 \\  
DDO\,210   & -10.9 &  0.45    & -1.9 &   2 & 1 \\
Leo\,I     & -11.9 &  1.2     & -1.4 &   2 & 1 \\
Peg\,DIG   & -12.9 &  1.6     & -2.0 &   2 & 1 \\
Fornax     & -13.1 &  2.0     & -1.2 &   2 & 1 \\ 
IC\,1613   & -15.3 &  2.3     & -1.4 &   2 & 1 \\

And\,II    & -11.8 &  0.3     & -11.5&   1 & 9,10\\

And\,III   & -10.2 &  \nodata & -1.7 &   1 & 2 \\ 
And\,V     &  -9.1 &  \nodata & -1.9 &   1 & 2 \\
And\,VI    & -11.3 &  0       & -1.7 &   1 & 2 \\ 
And\,VII   & -12.0 &  0.47    & -1.5 &   1 & 2 \\
Cetus      & -10.1 &  0       & -1.7 &   1 & 2 \\
NGC\,147   & -15.1 &  2.17    & -1.1 &   3 & 2 \\

NGC\,185   & -15.6 &  2.19    & -0.8 &   3 & 3 \\
Sag\,DIG   & -12.0 &  1.2     & -2.3 &   2 & 4 \\
NGC\,205   & -16.4 &  2.7     & -0.5 &   3 & 5 \\
Sgr        & -15.0 &  1.41    & -0.5 &   1 & 6 \\
SMC        & -17.1 &  3.35    & -1.2 &   2 & 7 \\
LMC        & -18.5 &  4.04    & -0.6 &   2 & 7 \\
M\,32      & -16.5 &  \nodata & -1.1 &   3 & 8 \\

\enddata

\tablenotetext{1}{Taken from \citet{grebel2003};
  Type: 1 - dSph, 2 - dIrr/transition type/dSph with recent star
  formation, 3 - dE}

\tablenotetext{2}{1 - \citet{battinelli2000}; 
                  2 - this work; 
                  3 - \citet{nowotny2003};
                  4 - \citet{demers2002};
                  5 - \citet{demers2003};
                  6 - \citet{whitelock1999};                  ;
                  7 - \citet{cioni2003};
                  8 - \citet{davidge2000};
	          9 - \citet{aaronson1985};
                 10 - \citet{cote1999}}

\label{tab_cstats}
\end{deluxetable}

\begin{deluxetable}{lccc}
\tablewidth{0pt}
\tablecaption{On the dC star frequency}

\tablehead{
 \colhead{Galaxy} &
 \colhead{\#RGB stars} &
 \colhead{\#dC stars} &
 \colhead{$\frac{ \rm{\#RGB stars}}{\rm{\#dC stars}}$ [\%]}
}

\startdata
\andiii &  439 & 1 & 0.23 \\
\andv   &  187 & 0 & 0.00 \\
\andvi  &  228 & 1 & 0.44 \\
\andvii & 1247 & 2 & 0.16 \\
Cetus   & 319  & 2 & 0.62 \\ \hline
All     & 2420 & 6 & 0.25
\enddata

\label{tab_dcfrequency}
\end{deluxetable}

\end{document}